\documentclass{vgtc}                          

\ifpdf
  \pdfoutput=1\relax                   
  \usepackage{graphicx}                
  \DeclareGraphicsExtensions{.pdf,.png,.jpg,.jpeg} 
\else
  \usepackage{graphicx}                
  \DeclareGraphicsExtensions{.eps}     
\fi%

\graphicspath{{figures/}{pictures/}{images/}{./}} 

\usepackage{microtype}                 
\PassOptionsToPackage{warn}{textcomp}  
\usepackage{textcomp}                  
\usepackage{mathptmx}                  
\usepackage{times}                     
\usepackage{cite}                      
\usepackage{tabu}                      
\usepackage{booktabs}                  
\usepackage{adjustbox}
\usepackage{subfig}
\usepackage{gensymb}

\onlineid{0}

\vgtccategory{IEEE VR 2020 Poster}

\vgtcinsertpkg

\title{Audio-Visual Spatial Alignment Requirements of Central and Peripheral Object Events}

\author{
Davide Berghi\thanks{Work supported by Polymersive: Immersive Video Production for Studio and Live Events, funded by Innovate UK (105168) with BBC R\&D and IMSRVRay; DoubleMe; Doctoral College, University of Surrey, UK. e-mails: d.berghi@surrey.ac.uk, hanne.stenzel@iis.fraunhofer.de, marco.volino@surrey.ac.uk, a.hilton@surrey.ac.uk, p.jackson@surrey.ac.uk}\\ %
  \scriptsize CVSSP, Univ. Surrey, UK %
\and Hanne Stenzel\\ %
  \scriptsize Fraunhofer IIS, Germany %
\and Marco Volino\\ %
  \scriptsize CVSSP %
\and Adrian Hilton\\ %
  \scriptsize CVSSP %
\and Philip J.B. Jackson\\ %
  \scriptsize CVSSP 
}

\abstract{Immersive audio-visual perception relies on the spatial integration of both auditory and visual information which are heterogeneous sensing modalities with different fields of reception and spatial resolution. This study investigates the perceived coherence of audio-visual object events presented either centrally or peripherally with horizontally aligned/misaligned sound. Various object events were selected to represent three acoustic feature classes. Subjective test results in a simulated virtual environment from 18 participants 
indicate a wider capture region in the periphery, with an outward bias favoring more lateral sounds.
Centered stimulus results support previous findings for simpler scenes.
} 

\CCScatlist{
  \CCScatTwelve{Virtual reality}{Audio-visual alignment}{Ventriloquism effect}{Peripheral stimuli};
}

\begin{document}

\firstsection{Introduction}

\maketitle

As the human brain senses the external world, it integrates signals belonging to the same object as a single percept. This is known as multimodal integration. For example, in a conversation scenario, one does not consider the talker's voice and lip motion as two distinct events. Instead, they are treated as a unified action: speech. 
This psychological fusion process allows for a certain tolerance in the timing and localization of the scene's object events. 
The tolerance introduced by the multimodal integration plays a key role in virtual reality and multimedia applications since object event signals are rarely perfectly aligned across all modalities. In the vast majority of entertainment, what the user sees and hears provides the main sensory inputs. So, in this paper, we investigate questions concerning the bimodal integration of audio and video. 
We research the spatial alignment requirements and extend previous literature to include peripheral positions using realistic audio-visual object events embedded within a simulated 3D environment. 

\section{Background: The Ventriloquism Effect}

The human brain tends to perceive audio and visual signals as a unified object even in the presence of a spatial mismatch between the locations of their sources, as long as the spatial misalignment is small enough. This illusion is known as the ventriloquism effect (VE) \cite{Alais:2004:VE}. 
The strength of the VE is influenced by several factors, such as the unimodal localization precision or the typology of audio-visual stimuli.
Human audio-visual localization has proven to behave differently when the stimuli are presented peripherally \cite{BongLee:2015:APV}. 
Stenzel \textit{et al.} \cite{Stenzel:2018:PTC} observed significant variations comparing stimuli with different acoustic features. Additionally, studies by Kyt{\"o} \textit{et al.} \cite{Kyto:2015:VEV} found that the VE could extend even farther when test participants are immersed in an AR scenario. 
Komiyama \cite{Komiyama:1989} and Stenzel \textit{et al.} \cite{Stenzel:2018:PTC} found that the participant's musical preparation influences one's ability to detect the spatial misalignment between audio and visual stimuli:  with musically untrained participants, the size of the VE approaches double that achieved with trained participants.
Discrete object events in VR applications are not only presented straight ahead but may also appear in peripheral areas. The current research is motivated by the need to define the differences in the VE between central and peripheral stimuli presentation, taking into account the variations induced by the type of stimulus and the participant training.
The visual stimuli employed in the tests are a 3D reconstruction of the items utilized by Stenzel \textit{et al.} \cite{Stenzel:2018:PTC} and were presented while the participants were immersed within the projection of a wide curved virtual environment.

\section{Methods}

The methods presented aim to assess the VE along the azimuth direction. A yes-no forced-choice 
test was adopted to assess whether the presented audio-visual spatial offset was perceivable or not. The data collected from the subjective tests have been interpolated using the psychometric function (PF) \cite{Wichmann:2001:PFF}, which relates the strength of a stimulus to the probability of its correct classification. The offset angle at which 50\% of the responses agree on classifying the stimuli as coherent is called the point of subjective equality (PSE), i.e the strength of the VE. 
The PF typically takes the shape of a Sigmoid function, normalized in the range [0,1], where 1 represents the absence of the tested attribute, i.e. when the stimuli are spatially aligned.
The PF proposed by Wichmann \textit{et al.} \cite{Wichmann:2001:PFF} is given by 
\begin{equation}
\Psi(x;\alpha,\beta,\gamma,\lambda) = \gamma + (1-\gamma-\lambda)\frac{1}{1+ \mathrm{exp}\big(-\beta(x-\alpha)\big)}
\end{equation}
where $x$ is the stimulus strength, $\alpha$ the overall curve's position, $\beta$ the slope, $\gamma$ \ the \textit{guess rate} which represents the lower bound of the curve, and the $\lambda$ the \textit{lapse rate}, i.e. the responses given regardless of the stimulus intensity.


\textbf{Dataset.} The dataset employed in the experiment is made up of 9 audio-visual stimuli. Each audio-visual stimulus consists of an audio clip and a volumetric video sequence made up of 3D geometry and a UV texture atlas at each time instance. 
Volumetric video sequences were captured in a multi-camera studio comprised of a 16-cameras set up in an inward facing 360\degree\ configuration. 
The dataset was partitioned into audio feature classes: ``Continuous" sounds, ``Harmonic" sounds, and ``Discrete" sounds, in order to study 
how different acoustic properties influence the VE.

\begin{figure}[tb]
 \centering 
 \includegraphics[width=0.80\columnwidth]{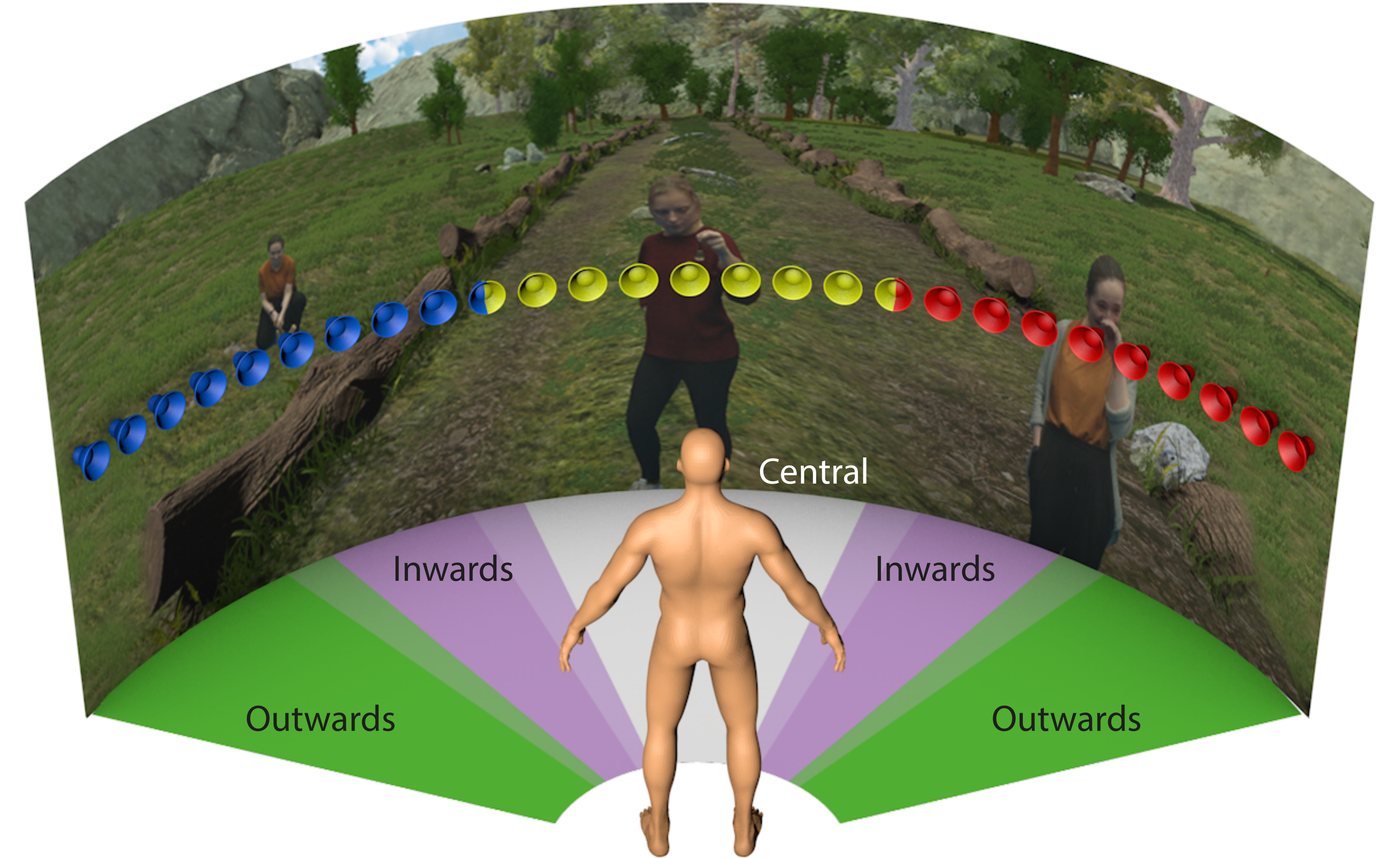}
 \caption{Loudspeaker positions overlaid on the projected virtual environment. Blue, yellow and red colors denote loudspeakers employed in audio stimulus reproduction.
 Division of the field of vision into central, inward and outward peripheral areas for analysis.}
 \label{fig:PIO}
\end{figure}

\textbf{Experimental Design.} With the aid of two projectors, the visual stimuli were projected on a white, curved and acoustically transparent screen covering approximately 150\degree\ in azimuth of the \textit{Surrey Sound Sphere}. 
During the experiment, the VR scenario was constantly projected, and the 3D objects were presented randomly either at 0\degree, +41.2\degree\ or -41.2\degree\ of the participants' 
head-forward direction.
The related audio signal was played time-synchronously with the visual object from one of the loudspeakers located behind the screen in the neighborhood of the visual position, as highlighted in \autoref{fig:PIO}.
The experiment was composed of a total of 288 stimuli presentations. 
The 3D objects were located within the virtual environment in order to 
be aligned with the three zero-offset loudspeakers.
Before each stimulus presentation, a circular target was projected in the center of the screen in order to focus the participant's sight centrally.
Participants were asked to avoid head movements,
yet they were permitted to re-direct their gaze toward the foreground visual stimulus.



\begin{table}[tb]
\centering
\resizebox{\columnwidth}{!}{%
\begin{tabular}{cccc|ccc|ccc}

& \multicolumn{3}{c}{ \textbf{Central}}&\multicolumn{3}{c}{ \textbf{Inward}}&\multicolumn{3}{c}{ \textbf{Outward}}\\
\hline
&	$\alpha$ & $\beta$ & PSE & $\alpha$ & $\beta$ & PSE & $\alpha$ & $\beta$& PSE \\ 
\hline

\textbf{C}  &0.59 &7.2 &9.0\degree& 0.55 & 6.6 & 9.7\degree & 0.58 & 7.2 & 11.4\degree\\

\textbf{D} &0.52  &5.0 & 10.6\degree& 0.55 & 4.6 & 10.2\degree & 0.42 & 5.2 & 15.7\degree\\
\textbf{H} &0.52 & 5.2 & 10.5\degree & 0.50 & 5.4 & 11.1\degree & 0.50 & 5.1 & 13.8\degree\\
\hline
\textbf{All} &0.55 & 5.7 & 10.0\degree & 0.53 & 5.6 & 10.4\degree & 0.52 & 5.8 & 13.2\degree\\
\hline
\end{tabular}}
\caption{Estimated PF parameters and PSEs for Discrete (D), Continuous (C), and Harmonic (H) audio feature classes.}
\label{tab:AFC}
\end{table}

\section{Results}

The parameters $\alpha$ and $\beta$ were estimated for different combinations of participant, audio-visual position, and stimulus levels. The audio-visual domain was divided into ``central", ``inward", and ``outward" positions as highlighted in the bottom part of \autoref{fig:PIO}.
In a final step, a Gaussian interpolation of the peripheral responses was performed to estimate the perceived coincidence angles. 

\textbf{Visual position.} To determine the effect of the visual positions on the PSEs, an analysis of variance (ANOVA) was conducted. The results show that the visual position and the participants have a significant effect on the PSEs but their interaction was not significant ($p>0.1$). A post-hoc comparison using the Tukey-HSD test indicated significant differences between the outward position and both inward ($p=0.004$) and central positions ($p=0.004$). These results are mirrored in the overall PSEs, with a PSE of 10.0\degree\ for the central position, 10.4\degree\ for the inward position, and 13.2\degree\ for outward audio-visual offsets.

\textbf{Audio feature classes.} Secondly, the PSEs were estimated for each item and visual position combination. The PSEs measured in the central field of vision range from 7.8\degree\ to 13.0\degree, for peripheral-inward offsets from 7.8\degree\ to 13.3\degree, while for outwards offsets from 10.4\degree\ to 18.5\degree. The ANOVA analysis revealed that both visual position and audio feature class influence the PSEs significantly. 
A further Tukey-HSD post-hoc analysis on the audio feature classes revealed a significant distinction between the Continuous sounds and the other two groups ($p=0.01$ for both comparisons).

\textbf{Coincidence angle.} A Gaussian interpolation outlines an overall shift of 1.4\degree\ outwards with respect to the visual stimulus position. The size of the outward shift varies across the items, up to a maximum of 3.3\degree. The Discrete sounds class produced the greatest shift (3\degree), whereas it was smallest for Continuous sounds (0.8\degree).

\textbf{Trained vs. Untrained.} The PF was re-estimated separately for trained and untrained participants per visual position. Results show that the PSEs increase by 60\%-100\% for untrained participants for each position group. The Gaussian interpolations revealed an outward shift in the coincidence angle of 0.9\degree\ and 2.6\degree\ for musically trained and untrained participants respectively.

\begin{figure}[tb]
\centering
\subfloat[][Central PFs]{
\includegraphics[width=0.28\columnwidth]{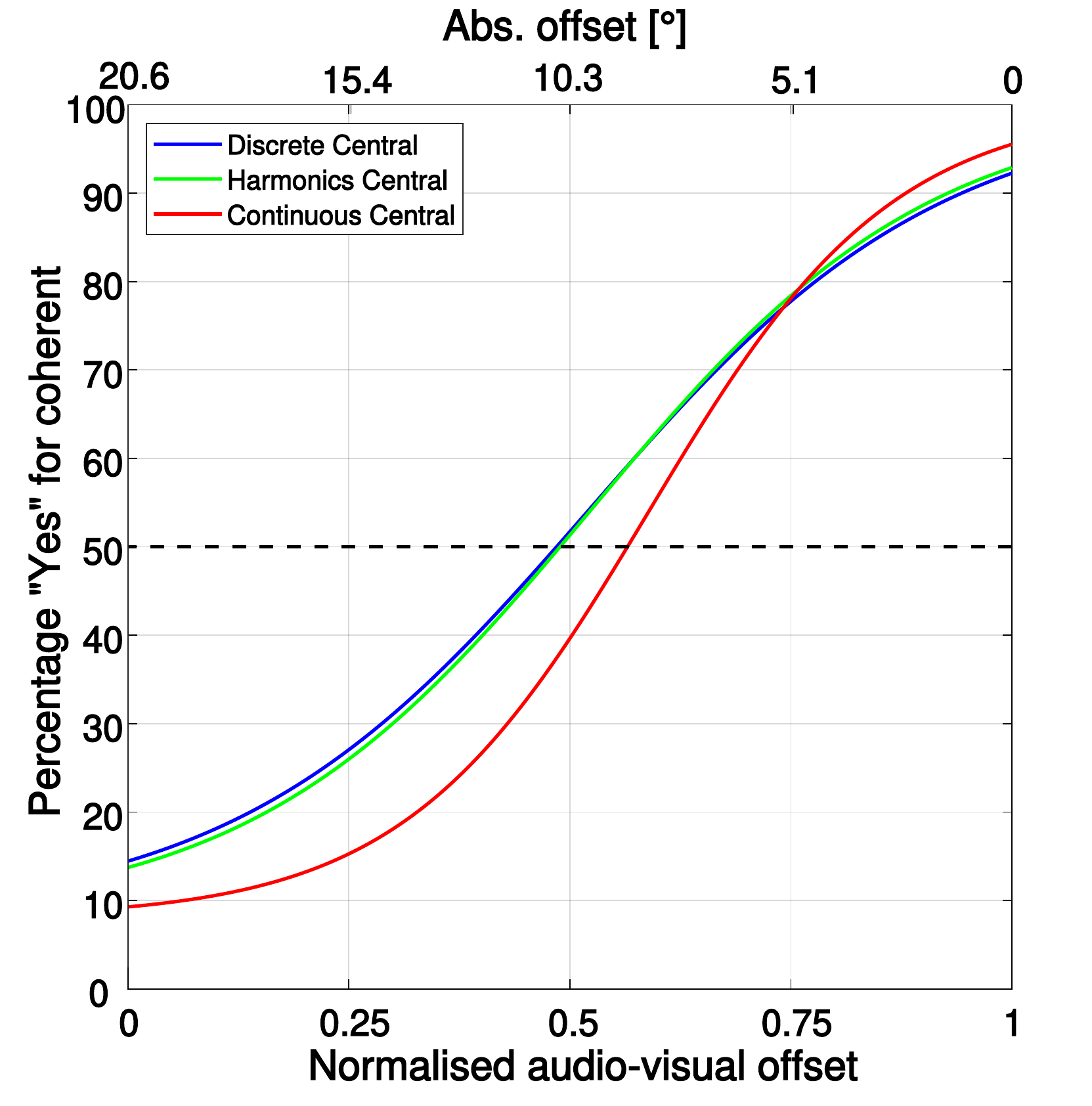} 
}
~
\subfloat[][Inward PFs]{
\includegraphics[width=0.28\columnwidth]{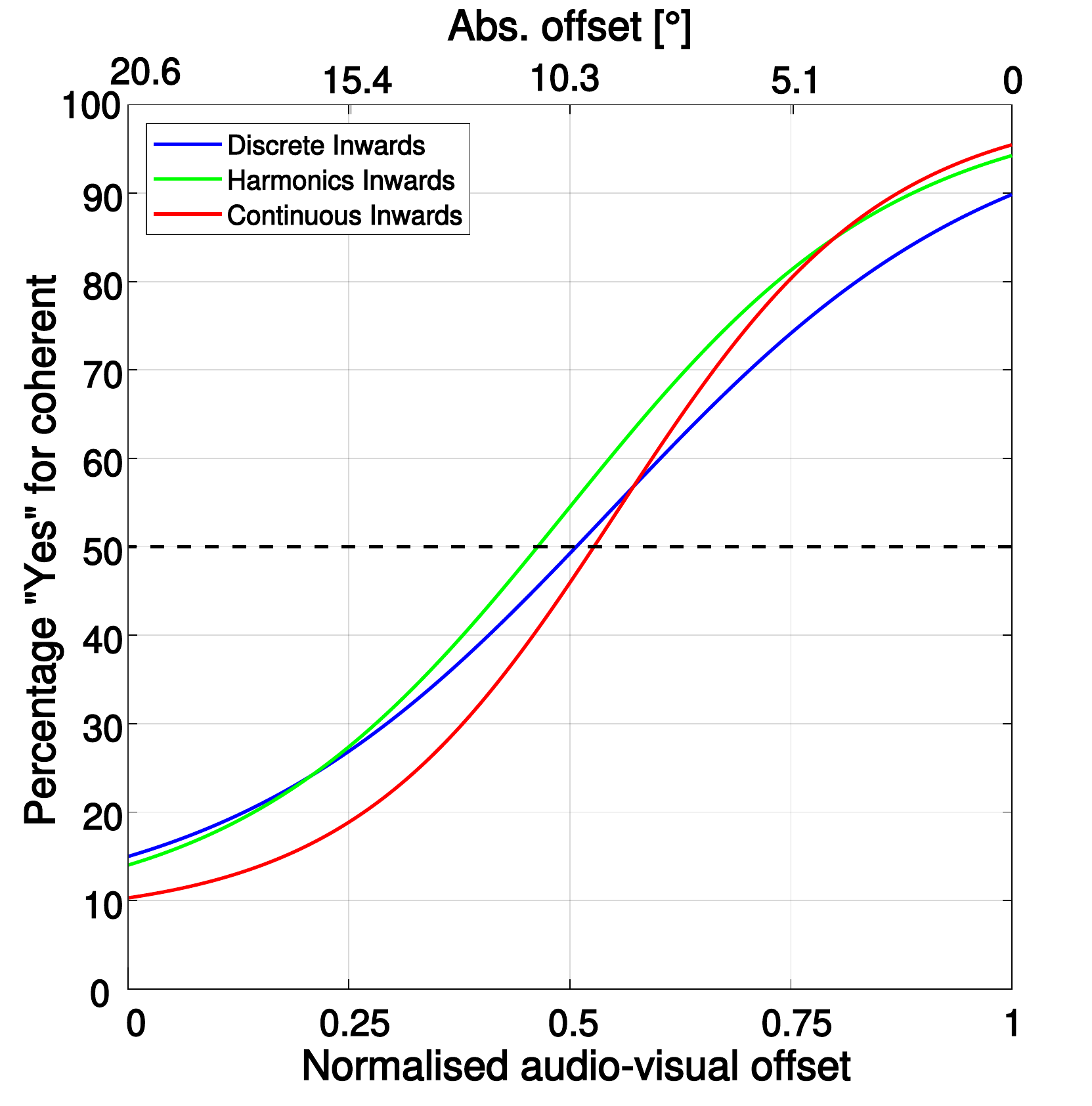} 
}
~
\subfloat[][Outward PFs]{
\includegraphics[width=0.28\columnwidth]{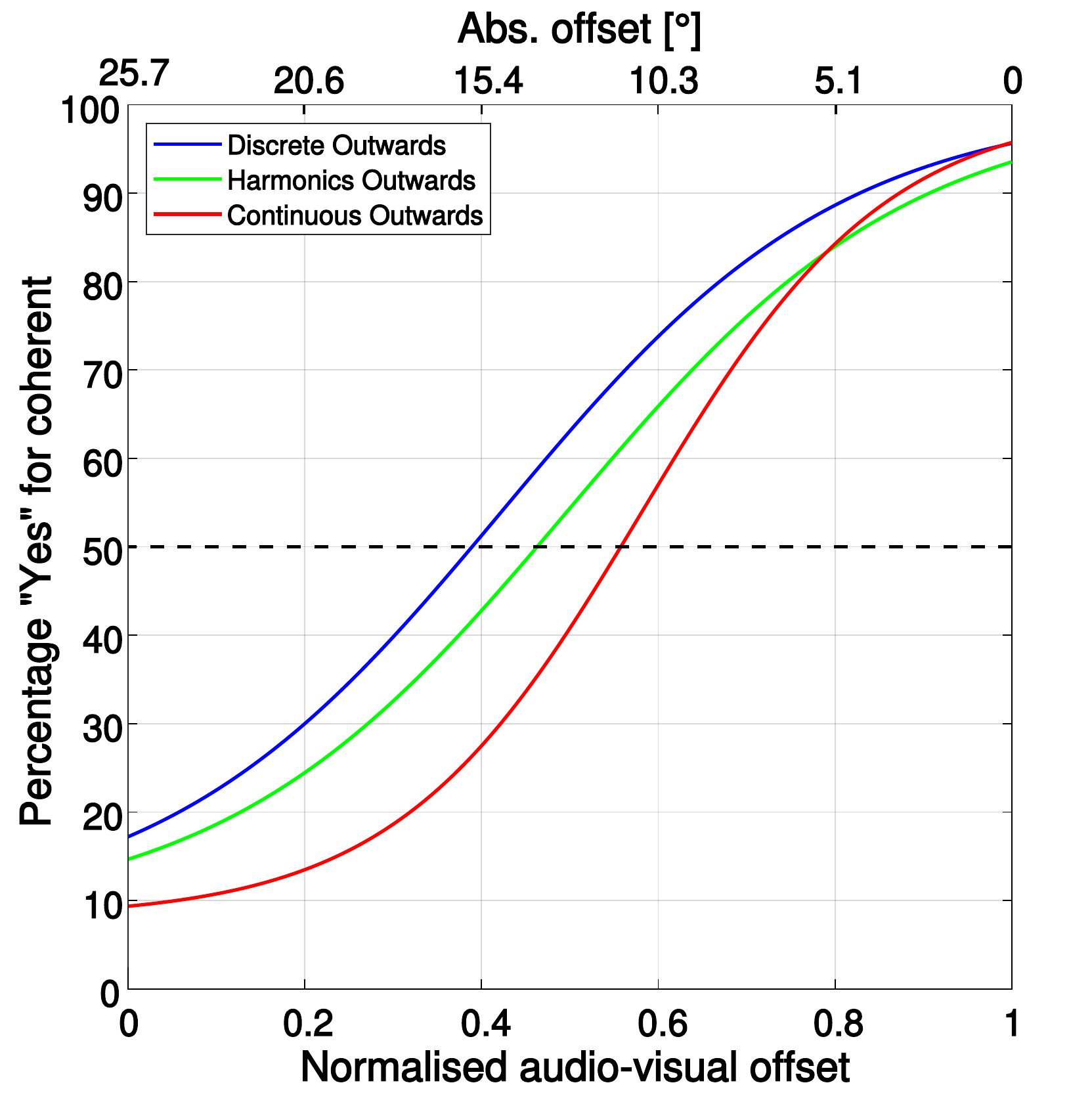} 
}

\subfloat[][Coincidence angles]{
\includegraphics[width=0.29\columnwidth]{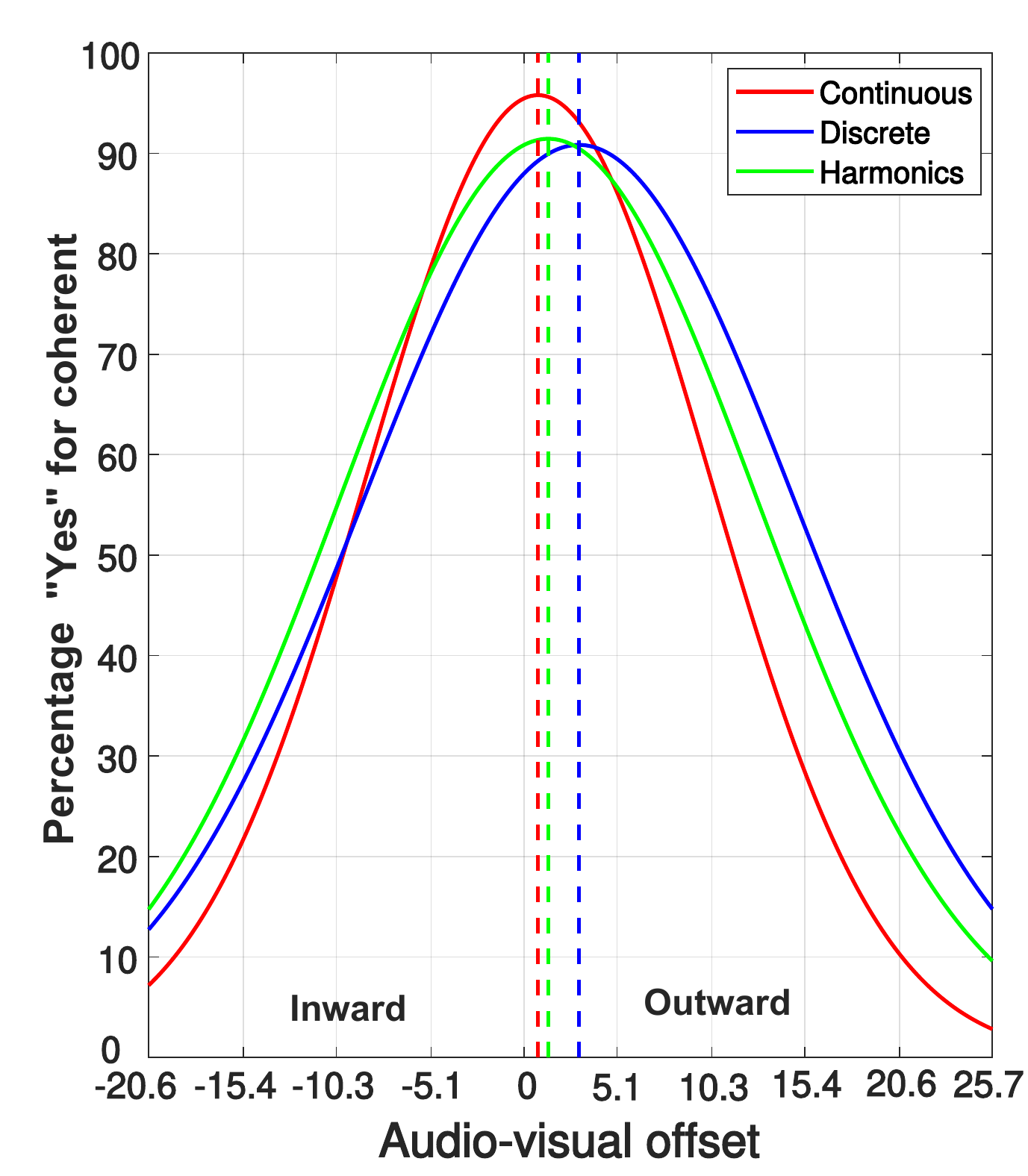} 
}
~
\subfloat[][Loudspeakers configuration]{
\includegraphics[width=0.60\columnwidth]{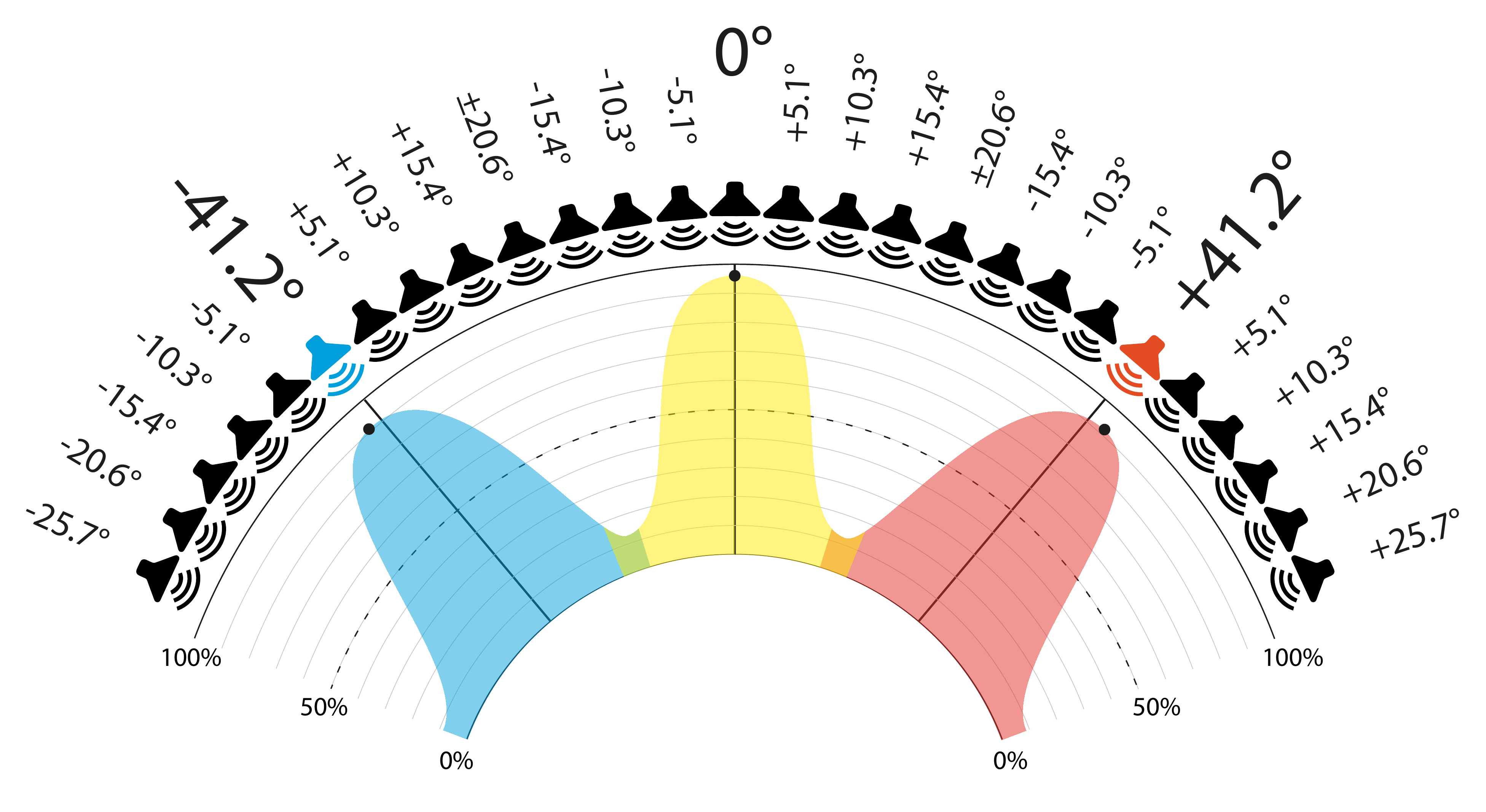} 
}
\caption{The PFs estimated per audio feature class AFC, for (a) the central position, (b) the inward position, and (c) the outward position. (d) the Gaussian interpolation per AFC and (e) the loudspeakers set up with overall subjective responses.}
\label{fig:GI}
\end{figure}

\section{Discussion \& Conclusion}

It was shown that the size of the VE increased at peripheral presentations and was significantly larger outward in the periphery.
The mean central PSE occurred at $\pm$10\degree\ offset.
In the periphery, the inward offset at PSE was slightly larger, whereas the outward offset increased to 13\degree. 
Such an increment is reflected in an outward shift of the perceived coincidence angle. 
Continuous sounds produced the smallest PSEs in both central and peripheral stimuli; 
Discrete sounds resulted in the greatest shift.
In all positions, ventriloquism had a stronger overall effect on untrained participants (15\degree) than trained (9\degree).
This effect was less marked for the inward periphery, consistent with a greater lateral bias of the perceived auditory location inferred from the coincidence angles: (visual 41.2\degree) trained 42.1\degree, untrained 43.8\degree.
Further tests can study more complex ecological scenes.


\bibliographystyle{abbrv-doi}

\bibliography{IEEEVR20_POSTER}

\end{document}